\documentclass[preprint]{aastex}

\begin{document}
\title{Discovery of the Bright Trans-Neptunian Object 2000 EB173}

\author{
Ignacio Ferrin\altaffilmark{1},
D. Rabinowitz	\altaffilmark{2},
B. Schaefer	\altaffilmark{2,3},
J. Snyder	\altaffilmark{2,3},
N. Ellman	\altaffilmark{2},
B. Vicente	\altaffilmark{4,8},
A. Rengstorf	\altaffilmark{5},
D. Depoy	\altaffilmark{6},
S. Salim	\altaffilmark{6},
P. Andrews	\altaffilmark{2},
C. Bailyn	\altaffilmark{3},
C. Baltay	\altaffilmark{2},
C. Briceno	\altaffilmark{2,4},
P. Coppi	\altaffilmark{2,3},
M. Deng		\altaffilmark{2},
W. Emmet	\altaffilmark{2},
A. Oemler	\altaffilmark{7},
C. Sabbey	\altaffilmark{9},
J. Shin		\altaffilmark{2},
S. Sofia	\altaffilmark{3},
W. van Altena	\altaffilmark{3},
K. Vivas	\altaffilmark{3},
C. Abad		\altaffilmark{4},
A. Bongiovanni	\altaffilmark{4},
G. Bruzual	\altaffilmark{4},
F. Della Prugna	\altaffilmark{4},
D. Herrera	\altaffilmark{4},
G. Magris	\altaffilmark{4},
J. Mateu	\altaffilmark{4},
R. Pacheco	\altaffilmark{4},
Ge. S\'{a}nchez	\altaffilmark{4},
Gu. S\'{a}nchez	\altaffilmark{4},
H. Schenner	\altaffilmark{4},
J. Stock	\altaffilmark{4},
K. Vieira	\altaffilmark{4},
F. Fuenmayor	\altaffilmark{1},
J. Hernandez	\altaffilmark{1},
O. Naranjo	\altaffilmark{1},
P. Rosenzweig	\altaffilmark{1},
C. Secco	\altaffilmark{1},
G. Spavieri	\altaffilmark{1},
M. Gebhard	\altaffilmark{5},
H. Honeycutt	\altaffilmark{5},
S. Mufson	\altaffilmark{5},
J. Musser	\altaffilmark{5},
S. Pravdo	\altaffilmark{10},
E. Helin	\altaffilmark{10},
K. Lawrence	\altaffilmark{10}
}

\altaffiltext{1}{Universidad de los Andes, Departamento de Fisica, 5101, 
M\'{e}rida, Venezuela, ferrin@ciens.ula.ve} 

\altaffiltext{2}{Yale University, Physics Department, P. O. Box 208121, 
New Haven CT 06520-8121}

\altaffiltext{3}{Yale University, Astronomy Department, P. O. Box 208101, 
New Haven CT 06520-8101}

\altaffiltext{4}{Centro de Investigaciones de Astronom\'{\i}a (CIDA), 
A. P. 264, 5101-A, M\'{e}rida, Venezuela}

\altaffiltext{5}{Indiana University, Deptartment of Astronomy, 319 Swain West, 
Bloomington IN 47405}

\altaffiltext{6}{Ohio State University, Department of Astronomy, 
Columbus OH 43210}

\altaffiltext{7}{Carnegie Observatories, 813 Santa Barbara St., 
Pasadena CA 91101}

\altaffiltext{8}{Universidad de Zaragoza, Grupo de Mecanica Espacial, 
Zaragoza, Spain 50009}

\altaffiltext{9}{Institute of Astronomy, Madingley Road, 
Cambridge, CB3 OHA, England}

\altaffiltext{10}{Jet Propulsion Laboratory, California Institute of Technology,
Pasadena CA 91109}

\slugcomment{
submitted to ApJ Letters 2000  Oct 21, 
revised 2000 Nov 28
}

\begin{abstract}

We describe the discovery circumstances and photometric properties of 2000
EB173, now one of the brightest trans-Neptunian objects (TNOs) with opposition
magnitude $m_{R}=18.9$ and also one of the largest Plutinos, found with the 
drift-scanning camera of the QUEST Collaboration, attached to the 1-m Schmidt 
telescope of the National Observatory of Venezuela.  
We measure  $B-V=0.99 \pm 0.14$ and
$V-R=0.57 \pm 0.05$, a red color observed for many fainter TNOs.    
At our magnitude limit $m_{R}=20.1 \pm 0.20$, our single detection reveals 
a sky density of 0.015 (+0.034, -0.012) TNOs per deg$^{2}$ (the
error bars are $68\%$ confidence limits), consistent with fainter
surveys showing a cumulative number proportional to
$10^{ 0.5 m_{R} }$.  Assuming an inclination distribution of TNOs with 
FWHM exceeding 30 deg, it is likely that one hundred to 
several hundred objects brighter than $m_{R}=20.1$ remain to 
be discovered.

\end{abstract}


\section{Introduction}

In recent years, the existence of a belt of small planets with orbits
beyond  Neptune has been firmly established following the discovery of
1992 QB1 (Jewitt \& Luu 1993). These  bodies are believed to be 
icy planetesimals remaining from the time preceding planet formation, 
as predicted by Edgeworth (1949) and Kuiper (1951). As of mid-2000, 
over 300 trans-Neptunian objects (TNOs)
are known (Marsden 2000), most having faint magnitudes
$(20< m_{R} <28)$ owing to their detection in narrow-field surveys with
large-aperture telescopes. Physical studies have been limited to
broadband filter photometry for the fainter bodies and spectroscopy
for a few of the brighter objects. These studies show that
TNOs are spectrally diverse, with colors ranging from neutral
to extremely red (Luu \& Jewitt 1996; Tegler \&
Romanishin 1998; Jewitt \& Luu 1998; Noll, Luu, \& Gilmore 2000;
Barucci  et al. 2000)
and with absorption features indicative of
hydrocarbons and water ice (Brown et al. 1997; Brown, Cruikshank, \&
Pendleton 1999). Previous
analyses of the discovery data have shown that the
number brighter than $m_{R}$ increases as $10^{ \alpha m_{R}}$, with
$ \alpha $ ranging from $0.52 \pm 0.02$ to $0.76 \pm 0.10$ (Jewitt,
Luu \& Trujillo 1998; Gladman et al. 1998; Chiang \& Brown 1999)
depending on the choice and interpretation of
discovery data. It is not known if this luminosity function
extends to magnitudes as bright as Pluto ($m_{V}=14$) or
instead cuts off at intermediate magnitudes. Such a cutoff
might reveal a termination in the growth phase of the
TNOs at the time of Neptune's formation (Bailey 1993;
Stern \& Colwell 1997).

Here we describe the discovery circumstances
and photometric properties of 2000 EB173, now one of the brightest
TNOs with opposition magnitude $ m_{R}=18.9$ and also one of 
the largest Plutinos
(a TNO with a stable orbit like Pluto completing two
orbits for every three by Neptune, thereby avoiding close
encounters with Neptune). We show that this detection
would not be likely unless the luminosity function
continues with a shallow slope ($ \alpha \sim 0.5$) at bright magnitudes,
$m_{R} < 19$.  We also show that the object's photometric properties
are similar to those of fainter TNOs with red color.

\section{Observations}

We discovered 2000 EB173 in a computer-aided search through digital
images, all recorded in a single 6-hour period on the night of 2000 March 15,
using the 1-m Schmidt telescope at Llano del Hato, Venezuela.  Our
collaboration, known as the Quasar Equatorial Survey Team (QUEST), has
designed and constructed for this telescope a digital camera consisting
of a 4 x 4 mosaic of charge-coupled devices (CCDs), each with 2048 x 2048
15-micron pixels (Snyder 1998). The system is designed for
drift-scanning.  With the telescope
position fixed, we align the parallel clocking direction of the CCDs with
the East-West motion of the drifting sky image. By clocking each CCD to
match the drift rate, we continuously scan 16 separate images. Each scan
is 0.588 deg wide in Declination (Dec) and a variable width in Right
Ascension (RA) that depends on the duration of the scan (up to 165 deg
on long clear nights). Because the columns in our 4 x 4 mosaic run
East-West, each column yields 4 scans of the same sky area separated in
time by  3.8 minutes (the drift time between CCDs). By providing each
North-South row in the
mosaic with a separate wide-band filter (Johnson U, B, V, or R), each
CCD in a column scans the sky through a different band-pass.

Normally, we use this instrument to search for quasars, supernovae,
and other variable objects by repeatedly scanning the same area
on a nightly basis. To conduct our TNO search, however, we limited the
search to an area 28.3 deg wide in RA (centered at RA = 13.249 hours,
Dec = -1.08 deg), and scanned the area twice in one night with a
separation of 3.99 hours. The total area covered was 66.8 deg$^{2}$,
imaged four times per scan through the filter sequence V, R1, B, R2.
Later,
we processed the data by co-adding the two R and one V image from each
scan sequence. We then subtracted a scaled reference image created
from co-added scans of the same area recorded on previous nights. 
This removed images of stationary and non-varying stars and galaxies, leaving
only images of transient objects, asteroids, and noise artifacts.

We then used a computer program to identify objects with
integrated intensities in R1+R2+V greater than 3.0 times the noise in the
sky background.
By combining the set of objects
identified in both scan sequences and displaying their images
on a computer monitor, possible TNOs appeared as pairs of images.
The following constraints were then applied to each object pair: 
to show retrograde motion; to move within $ \pm 25$ deg of
the ecliptic; to have a separation of ~5 to 20 arcseconds (the
expected displacement in 3.99 hours of a TNO observed near 
opposition); to reveal
the same approximate magnitude in the 3.99-hour interval; to exhibit a FWHM
consistent with surrounding stars; and to be detected in the
separate R1, R2, and V images before co-addition (6 independent 
observations). The only candidate (out of ~600,000) to meet all these
requirements was 2000 EB173.

\section{Results}

{\em Efficiency}.  We estimated the limiting magnitude of our search 
by identifying main-belt asteroids appearing in a single scan sequence
taken near opposition on an earlier night through the same filter set.
Given the ~7.6-minute interval between successive R exposures in a 
single scan, the near-opposition motion of main-belt asteroids is 4 
to 6 pixels. We could therefore use the detection procedure 
described above, but with the two R images replacing each of the 
two co-added R+R+V images from the TNO scans.  

Figure 1 shows the number of main-belt asteroids we detect as a function 
of $m_{R}$. Our relative detection efficiency is the observed 
distribution divided by a 
power-law fit at bright magnitudes (Rabinowitz 1994). As shown
in the figure, our 50\% 
detection limit occurs at $m_{R}=19.6\pm0.2$. We also determined that 
our absolute detection efficiency at bright magnitudes is close to 100\% 
by verifying incidental detections of catalogued asteroids within the 
search fields.

We expect that our detection efficiency for TNOs has the same form as for 
main-belt asteroids, but shifted to fainter magnitudes because we 
co-add two R images and one V image for each TNO detection. This 
increases the relative signal from each TNO and from the sky by a factor 
of $2.6\pm0.1$, as verified from our measurements of 2000 EB173. The signal 
to noise ratio increases by the square root of this factor since the 
noise at the detection limit is dominated by counting statistics of the 
sky signal. Because we searched for TNOs and main-belt asteroids with 
the same threshold value of the signal to noise ratio, the limiting 
magnitude for our TNO search is fainter by a factor 
$2.5\log(\sqrt{2.6\pm0.1})$, yielding magnitude limit $m_{R}=20.1 \pm 0.2$.

{\em Orbit}.  Following the discovery of 2000 EB173, we confirmed its
motion in QUEST scans from previous nights. Based on an orbit solution
from these initial observations we recovered the object three months
later with the 1-m Yale telescope at Cerro Tololo, Chile.
Table 1 shows the most recent orbital elements determined for
2000 EB173 from observations spanning 4 years (Marsden and Williams
2000a). These include pre-discovery, photographic observations with the 48" 
Schmidt telescope at Palomar, and post-discovery observations by amateur 
astronomers with telescopes of sub-meter aperture (Marsden and Williams 2000b).  
The eccentricity, $e$, and inclination, $i$, are typical values
for a Plutino (Jewitt \& Luu 1996), while the semimajor axis, $a$, is relatively
low. Given the long arc of the observations, 
the orbit solution is accurate to better than $1\%$.

{\em Size}.  The absolute
magnitude we determine, $H=4.7 \pm 0.1$, is one of the brightest for any known
TNO, making 2000 EB173 also one of the largest TNOs assuming all
have similar albedos. The only brighter TNO is 1996 TO66 with $H=4.5$ and the
next brightest is 1999 TC36, a Plutino with $H=4.8$ (Marsden 2000). 
There are no known geometric albedos for TNOs, so one has to
be assumed.  For an albedo of 0.04, typical of dark,
carbon-rich asteroids and cometary nuclei and a value commonly assumed for TNOs, 
the corresponding
diameter would be $\sim600$ km, about 1/4 the size of Pluto (Jewitt,
Luu \& Trujillo 1998) and marginally resolvable with the Hubble Space
Telescope.

{\em Cumulative number}.  Figure 2 shows the number of TNOs per 
deg$^{2}$ we find brighter than our detection limit, plotted 
together with previous
measurements at mostly fainter limits. Also shown are two previously
suggested curves to fit the observations, $ \alpha = 0.58 \pm 0.05$ from
Jewitt, Luu \& Trujillo (1998) and $ \alpha = 0.76 \pm 0.1 $ from
Gladman et al. (1998). Our detection of 1
TNO per 66.8 deg$^{2}$ favors the $ \alpha =0.58 $ curve.
It also rules out the previously established upper limit (Kowal 1989) 
of ~1 TNO per 1000 deg$^{2}$ brighter than $m_{R} =19.5$, as this would
require an unrealistic, sharp cutoff in the magnitude frequency
above $m_{R}=20.1$. Given previously predicted magnitude-frequency
curves that assume an upper limit to the diameter distribution of
TNOs (Jewitt, Luu, \& Trujillo 1998), the likelihood of our discovery would be less
than $10 \%$ for any diameter cutoff below 500 km.

{\em Photometry and imaging}.  Following our discovery of 
2000 EB173 we made photometric R-band observations with the 1-m Yale telescope, 
BVRI observations with the 2.4-m Michigan-Dartmouth-MIT (MDM)
telescope at Kitt Peak, and V and R observations with the
3.5-m Wisconsin-Indiana-Yale-NOAO (WIYN) telescope, also at Kitt Peak.
Incidental, unfiltered CCD observations were also made by the Near Earth
Asteroid Tracking Program (NEAT) of the Jet Propulsion Laboratory using a 
U.S. Air Force 1.2-m telescope at Haleakala, Maui (these observations 
were identified using the SkyMorph archive at 
\url{http://skys.gsfc.nasa.gov/skymorph/skymorph.html}).

Table 2 lists the results from these observations and from
pre-discovery QUEST scans.  From Mar 1 to Mar 15, and from June 3 to
June 15, no variability in
$m_{R}$ greater than $10 \%$ appears. Furthermore, a series of
9 images in V taken June 15 shows no variability greater than
$ 3 \%$ over a 1.25-hour period. Co-addition of the June 15 images
reveals a stellar point-spread function for 2000 EB173, with any
cometary activity contributing less than $ 10 \%$ to the flux
outside a radius of 0.6 arcseconds. We are therefore confident that
the colors we measure are unaffected by intrinsic variability from
rotation or cometary outbursts.

{\em Colors}. Figure 3 shows the spectral reflectance we derive for 
2000 EB173 along with the measured reflectance for other TNOs 
(Jewitt \& Luu 1998).  Shortwards of 0.7 microns, the spectral 
slope we observe is consistent with that of other
other TNOs. This result places the object among a group with red
colors identified previously by Tegler \& Romanishin (1998). 
Longwards of 0.7 microns, we observe a possible transition to
a neutral reflectance, but this result is ambiguous. 
Depending on the pixel aperture we use for our photometry,
we also obtain a red $R-I$ color for 2000 EB173, which would be
expected based on previous observations of fainter red TNOs. 
In only a few cases has such a flattening of the reflectance 
been observed (e.g., 1993 FW as observed by Barucci et al. 2000).

\section{Conclusions}

1. The unusual brightness of 2000 EB173 provides an opportunity to study
the physical nature of one TNO with substantially greater precision
than has been possible previously. It now becomes possible for all astronomers,
including those with modest facilities, to make a valuable contribution to the study of
the primordial bodies preserved in the outer solar system.  Possible
contributions would be to determine the size of 2000 EB173 by measuring the albedo
with infrared observations or by direct observation with space-based telescopes.  
Such studies would have important implications for the total 
mass of the TNO population, currently uncertain by more than an order of magnitude.

2. We observe no variability in the lightcurve of 2000 EB173 and no cometary coma.
The color is very red shortwards of 0.7 microns, typical of many fainter TNOs.
Our measurements indicate a possible flattening of the reflectance
longwards of 0.7 microns, but are also consistent with a redder color. 

3. Our detection of one TNO brighter than $m_{R}=20.1$ in a search of 66.8 
deg$^{2}$ favors the flat luminosity function ($ \alpha = 0.58$) of 
Jewitt, Luu, \& Trujillo (1998) over the steeper distribution ($ \alpha = 0.768$)
of Gladman et al. (1998). 

4. Extrapolating to a search of the entire area within 60 deg of the 
ecliptic (within the inclination range for TNOs determined by Jewitt, Luu, \&
Chen 1996), 
one could reasonably expect to find from one hundred to several hundred TNOs 
brighter than $m_{R}=20.1$. A 
dedicated search using a telescope with  capabilities similar to our own would 
discover most of these bodies in a few years.

\acknowledgements

This work was supported by the National Science Foundation, the Department
of Energy, and the National Aeronautics and Space Administration. The
Observatorio Astron\'{o}mico Nacional is operated by
CIDA for the Consejo Nacional de Investigaciones Cient\'{\i}ficas y
Tecnol\'{o}gicas.  The Council for Scientific and Technological
Developmentof the University of the Andes also provided support.

Correspondence and requests for materials should be addressed to
David Rabinowitz, Yale University Physics Dept., P. O. Box 208121,
New Haven CT 06520-8121; or by E-mail to david.rabinowitz@yale.edu.

\clearpage
\begin{deluxetable}{cccccccc}  
\tablecolumns{8}  
\tablewidth{0pc}  
\tablecaption{Orbital Elements for 2000 EB173}  
\tablehead{  
\colhead{$a$(AU)} 
& \colhead{$e$} 
& \colhead{$i(^{\circ})$}   
& \colhead{$w(^{\circ})$}    
& \colhead{$Node(^{\circ})$} 
& \colhead{$M(^{\circ})$}    
& \colhead{$H$}  
& \colhead{Epoch} 
}   
\startdata  
39.003 & 0.266 & 15.493 & 66.981 & 169.359 & 339.028 & $4.7\pm0.1$ & 2451800.5 \\
\enddata  
\tablecomments{Orbital Elements are J2000 from Marsden \& Williams (2000a) with uncertainties
less than $1\%$. The value for $H$
derives from the R magnitude, $m_{R}=18.90\pm0.10$, observed
at the lowest phase angle, $\alpha=0.29^{\circ}$,
on 2000 April 11 UT (see Table 2) and from the mean $V-R$
color ($ 0.574 \pm 0.047$) determined from observations at all phase
angles.}
\end{deluxetable}  

\clearpage
\begin{deluxetable}{rrrrrrrrrr}  
\tablecolumns{10}  
\tablewidth{0pt}  
\tablecaption{Photometric Observations of 2000 EB173}  
\tablehead{  
\colhead{Date(2000)} 
& \colhead{$B$}   
& \colhead{$V$}    
& \colhead{$R$} 
&  \colhead{$I$}    
& \colhead{$B-V$}   
& \colhead{$V-R$}    
& \colhead{$V-I$}
& \colhead{$a(^{\circ})$}
& \colhead{Telescope}
} 
\startdata  
Feb 11.30 & \nodata & 20.01 & 19.32 & \nodata & \nodata & 0.69 & \nodata & 
1.60 & 1-m Schmidt \\
\nodata & \nodata & $\pm$0.17 & $\pm$0.06 & \nodata & \nodata & $\pm$0.18 & 
\nodata & \nodata & \nodata \\
Feb 12.30 & \nodata & 19.78 & 19.34 & \nodata & \nodata & 0.44 & \nodata & 
1.59 & 1-m Schmidt \\
\nodata & \nodata & $\pm$0.15 & $\pm$0.06 & \nodata & \nodata & $\pm$0.16 & 
\nodata & \nodata & \nodata \\
Feb 13.30 & \nodata & \nodata & 19.27 & \nodata & \nodata & \nodata & \nodata &
1.57 & 1-m Schmidt \\
\nodata & \nodata & \nodata & $ \pm 0.05$ & \nodata & \nodata & \nodata & \nodata & 
\nodata & \nodata \\
Feb 14.30 & \nodata & \nodata & 19.51 & \nodata & \nodata & \nodata & \nodata &
1.55 & 1-m Schmidt \\
\nodata & \nodata & \nodata & $ \pm 0.10$ & \nodata & \nodata & \nodata & \nodata & 
\nodata & \nodata \\
Mar 01.34 & \nodata & 19.84 & 19.30 & \nodata & \nodata & 0.54 & \nodata & 
1.19 & 1-m Schmidt \\
\nodata & \nodata & $\pm$ 0.14 & $\pm$0.05 & \nodata & \nodata & $\pm$0.15 & 
\nodata & \nodata & \nodata \\
Mar 03.33 & \nodata & 19.99 & 19.32 & \nodata & \nodata & 0.67 & \nodata & 
1.14 & 1-m Schmidt \\
\nodata & \nodata & $\pm$0.17 & $\pm$ 0.06 & \nodata & \nodata & $\pm$0.18 & 
\nodata & \nodata & \nodata \\
Mar 08.28 & \nodata & 20.03 & 19.27 & \nodata & \nodata & 0.76 & \nodata & 
1.01 & 1-m Schmidt \\
\nodata & \nodata & $ \pm 0.15$ & $ \pm 0.05$ & \nodata & \nodata & $ \pm 0.16$ & 
\nodata & \nodata & \nodata \\
Mar 10.30 & \nodata & 19.79 & 19.39 & \nodata & \nodata & 0.40 & \nodata & 0.95 & 
1-m Schmidt \\
\nodata & \nodata & $ \pm 0.16$ & $ \pm 0.08$ & \nodata & \nodata & $ \pm 0.18$ & 
\nodata & \nodata & \nodata \\
Mar 11.28 & \nodata & 19.80 & 19.19 & \nodata & \nodata & 0.61 & \nodata & 0.92 & 
1-m Schmidt \\*
\nodata & \nodata & $ \pm 0.14$ & $ \pm 0.06$ & \nodata & \nodata & $ \pm 0.015$ & 
\nodata & \nodata & \nodata \\*
Mar 15.19 & \nodata & 19.65 & 19.31 & \nodata & \nodata & 0.34 & \nodata & 0.81 & 
1-m Schmidt \\*
\nodata & \nodata & $ \pm 0.16$ & $ \pm 0.08$ & \nodata & \nodata & $ \pm 0.18$ & 
\nodata & \nodata & \nodata \\*
Mar 15.35 & \nodata & 19.94 & 19.21 & \nodata & \nodata & 0.73 & \nodata & 0.81 & 
1-m Schmidt \\*
\nodata & \nodata & $ \pm 0.15$ & $ \pm 0.05$ & \nodata & \nodata & $ \pm 0.16$ & 
\nodata & \nodata & \nodata \\
Apr 11.47 & \nodata & \nodata & 18.90 & \nodata & \nodata & 
\nodata & \nodata & 0.29 & NEAT 1.2-m \\
\nodata & \nodata & \nodata & $ \pm 0.10$ & \nodata & \nodata & 
\nodata & \nodata & \nodata & \nodata \\
Jun 03.99 & \nodata & \nodata & 19.40 & \nodata & \nodata & 
\nodata & \nodata & 1.62 & Yale 1-m \\
\nodata & \nodata & \nodata & $ \pm 0.12$ & \nodata & \nodata & 
\nodata & \nodata & \nodata & \nodata \\
Jun 10.16 & 21.10 & 20.11 & 19.51 & 19.13 & 0.99 & 
0.60 & 0.98 & 1.73 & MDM 2.4-m \\
\nodata & $\pm$0.11 & $\pm$0.08 & $\pm$0.06 & $\pm$0.07\tablenotemark{a} & $\pm$0.14 &
$\pm$0.10 & $\pm$0.11\tablenotemark{a} & \nodata & \nodata \\
Jun 15.20 & \nodata & 20.18 & 19.63 & \nodata & \nodata & 
0.55 & \nodata & 1.80 & WIYN 3.5-m \\
\nodata & \nodata & $ \pm 0.03$ & $ \pm 0.03$ & \nodata & \nodata & 
$ \pm 0.05$ & \nodata & \nodata & \nodata \\
\enddata 
\tablenotetext{a}{There is an additional systematic uncertainty of -0.22 in
$I$ and +0.22 in $V-I$.}
\tablecomments{A common set of calibrated field stars were photometric 
standards for the 1-m Schmidt observations. For the Yale and WIYN 
observations, the photometric standard was a single field star calibrated
against catalogue standards with MDM. The NEAT observations were CCD
observations with no filter, calibrated against secondary photometric 
standards observed in V an R with the 1-m Schmidt (a color correction term
was determined assuming the mean $V-R$ color, $0.574 \pm 0.047$, observed for
2000 EB173 on other nights). All colors are Kron-Cousins.}
\end{deluxetable}  


\begin{figure}
\begin{center}
\resizebox{10cm}{10cm}{\includegraphics{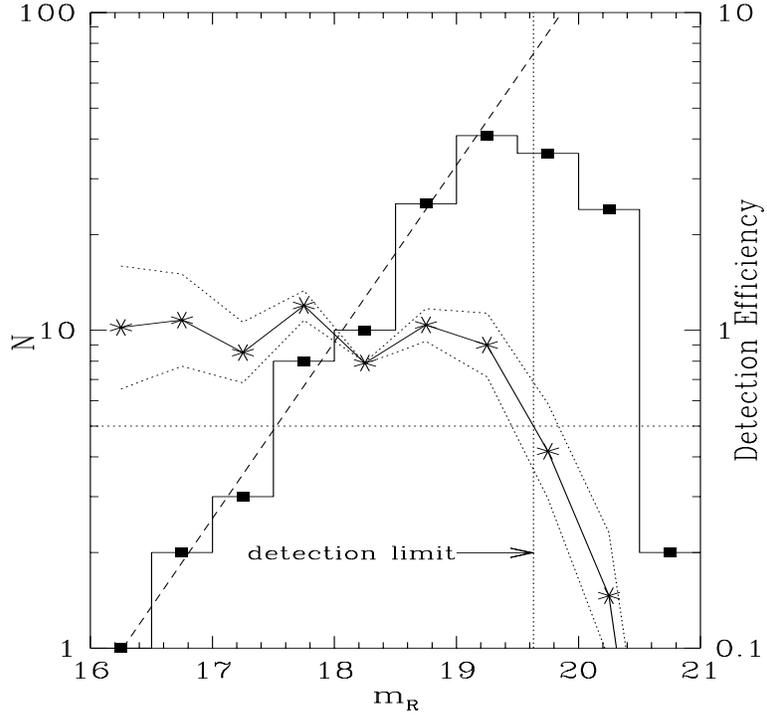}}
\caption{The number of detected main-belt asteroids, $N$ (squares), 
and the detection efficiency (stars) vs apparent R magnitude, 
$m_{R}$. The detection efficiency is $N$ divided by a power-law fit to $N$
vs $m_{R}$ for $m_{R}<19.0$ (dashed line). The dotted curves show
the uncertainty in the efficiency owing to the uncertainty in the
fit. The dotted lines show the 50\% detection limit at 
$m_{R}=19.63\pm0.20$. For TNOs, the detection limit is fainter by
0.51 magnitudes.
}
\end{center}
\end{figure}

\begin{figure}
\begin{center}
\resizebox{10cm}{10cm}{\includegraphics{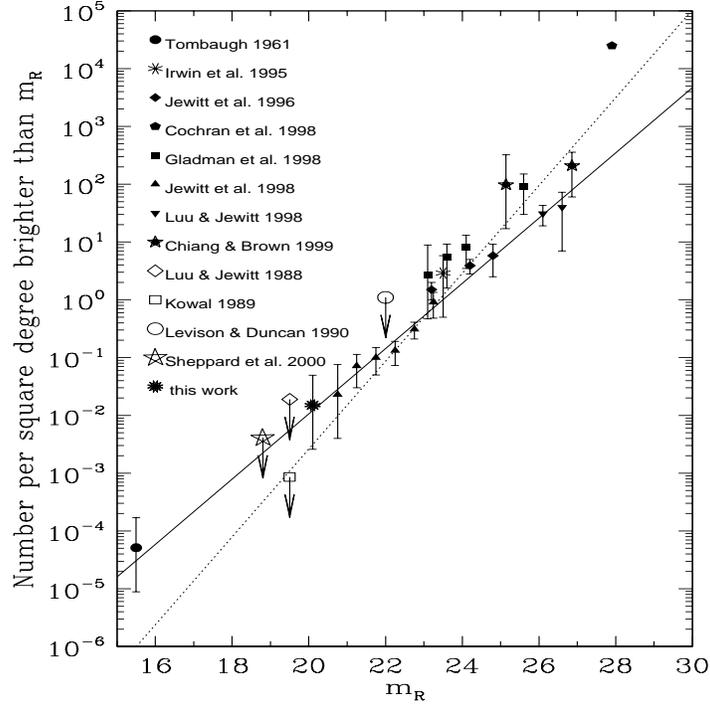}}
\caption{The cumulative number of TNOs per deg$^{2}$ brighter than
magnitude limit $m_{R}$ from this paper and from previous surveys. Where the
number per deg$^{2}$ is based on a single detection, we assign error
bars corresponding to a $68 \%$ confidence interval (i.e., the lowest and 
highest number per deg$^{2}$, $n$, for which the likelihood of detecting at 
least and at most one TNO, respectively, exceeds $16 \%$. For a given search 
area, $A$, this corresponds to the respective values of $n$
for which the summations from $k$ = 1 to $\infty$ and from $k$ = 0 to 1 of the 
Poisson likelihood $e^{-nA}(nA)^{k}/k!$ equal 0.16). Where the survey limit
is reported as a V magnitude (Luu \& Jewitt 1998, Kowal 1989, Cochran et
al. 1998), we assume $V-R=0.5$.  The solid
and dashed lines are two previously derived fits
(Jewitt, Luu, \& Trujillo 1998; Gladman et al. 1998) that assume a magnitude
frequency proportional to $10^{ \alpha m_{R} }$ with $ \alpha =
0.58 \pm 0.05$ and $ \alpha =0.76 \pm 0.11$, respectively.  
}
\end{center}
\end{figure}


\begin{figure}
\begin{center}
\resizebox{10cm}{10cm}{\includegraphics{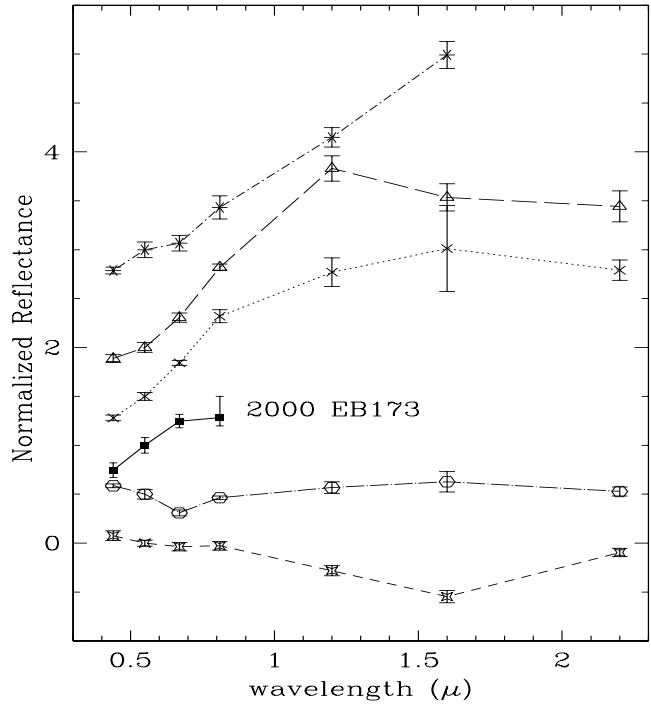}}
\caption{Spectral reflectance for 2000 EB173 (filled squares) and for
previously observed TNOs (from top to bottom: 1996 TS66, 1996 TP66, 1993 SC,
1996 TL66, and 1996 TO66) assuming solar colors ($B-V$,
$V-R$, and $V-I$) given by Jewitt \& Luu (1998).  All spectra are
normalized to unity at 0.55 microns.  They have been shifted vertically,
but share the same scale.  
}
\end{center}
\end{figure}

\end{document}